\documentclass[10pt,twocolumn,letterpaper]{article}

\usepackage[pagenumbers]{cvpr} 

\usepackage{graphicx}
\usepackage{amsmath}
\usepackage{amssymb}
\usepackage{booktabs}
\usepackage{bm}
\usepackage{multirow}
\usepackage{makecell}
\usepackage{caption}
\usepackage{stfloats}

\usepackage[pagebackref,breaklinks,colorlinks]{hyperref}

\usepackage[capitalize]{cleveref}
\crefname{section}{Sec.}{Secs.}
\Crefname{section}{Section}{Sections}
\Crefname{table}{Table}{Tables}
\crefname{table}{Tab.}{Tabs.}

\def\cvprPaperID{11239} 
\def\confName{CVPR}
\def\confYear{2022}

\begin{document}

\title{Hypernet-Ensemble Learning of Segmentation Probability for Medical Image Segmentation with Ambiguous Labels}

\author{Sungmin Hong\\
MGH, HMS 
\and
Anna Bonkhoff \\ 
MGH, HMS
\and
Andrew Hoopes \\
MGH, HMS
\and
Martin Bretzner \\
MGH, HMS
\and
Markus Schirmer \\
MGH, HMS
\and
Anne-Katrin Giese \\
UKE
\and
Adrian Dalca \\
MGH, HMS \& MIT
\and
Polina Golland \\
MIT CSAIL
\and
Natalia Rost \\
MGH, HMS 
}

\maketitle

\begin{abstract}
Despite the superior performance of Deep Learning (DL) on numerous segmentation tasks, the DL-based approaches are notoriously overconfident about their prediction with highly polarized label probability. 
This is often not desirable for many applications with the inherent label ambiguity even in human annotations. 
This challenge has been addressed by leveraging multiple annotations per image and the segmentation uncertainty. 
However, multiple per-image annotations are often not available in a real-world application and the uncertainty does not provide full control on segmentation results to users. 
In this paper, we propose novel methods to improve the segmentation probability estimation without sacrificing performance in a real-world scenario that we have only one ambiguous annotation per image.
We marginalize the estimated segmentation probability maps of networks that are encouraged to under-/over-segment with the varying Tversky loss without penalizing balanced segmentation.
Moreover, we propose a unified hypernetwork ensemble method to alleviate the computational burden of training multiple networks. 
Our approaches successfully estimated the segmentation probability maps that reflected the underlying structures and provided the intuitive control on segmentation for the challenging 3D medical image segmentation.
Although the main focus of our proposed methods is not to improve the binary segmentation performance, our approaches marginally outperformed the state-of-the-arts. The codes are available at \url{https://github.com/sh4174/HypernetEnsemble}.
\end{abstract}

\begin{figure}[t!]
  \centering
  \includegraphics[width=\linewidth]{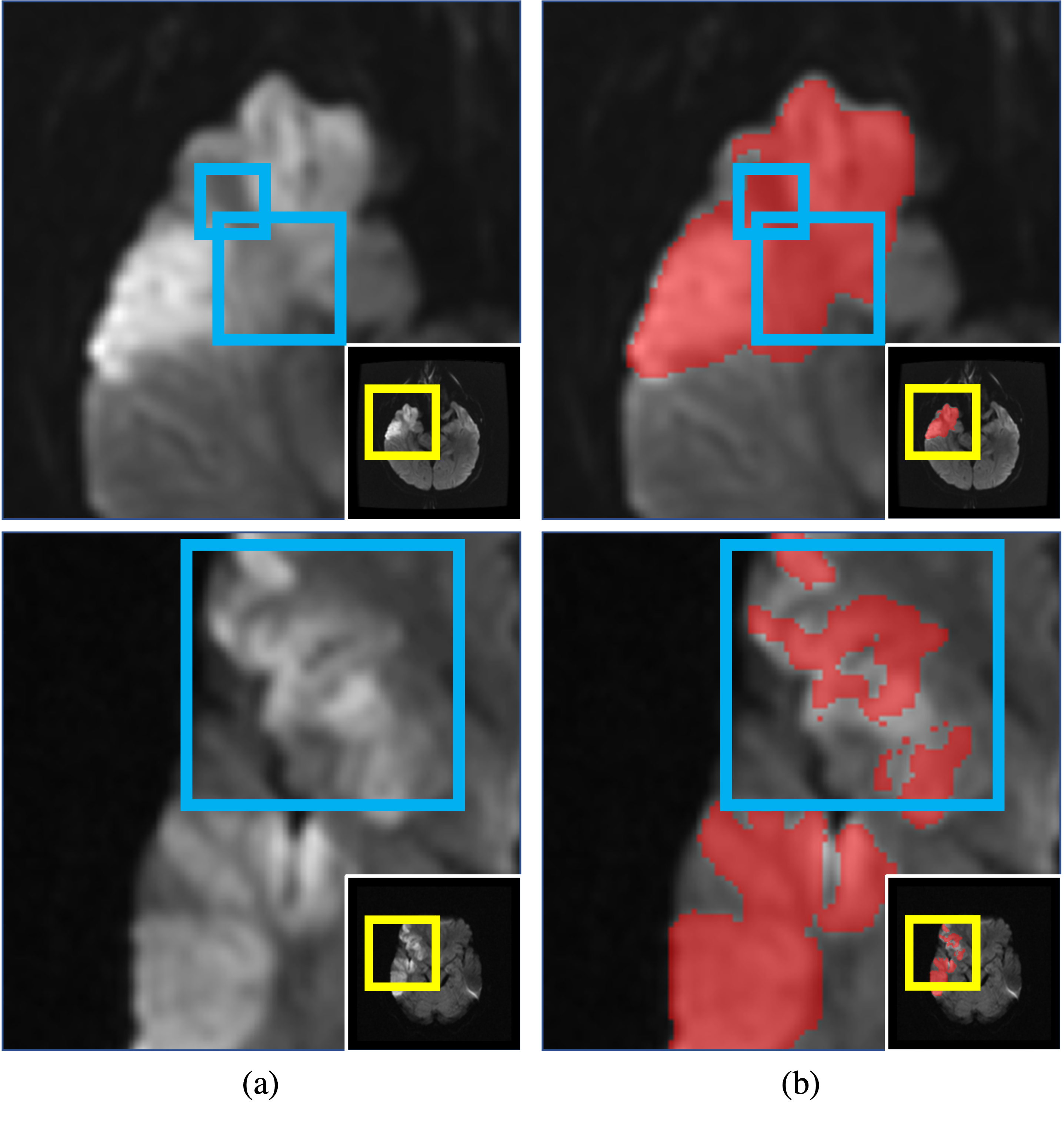}
  \vspace*{-8mm}
  \caption{Human annotations on ambiguous stroke lesions. Ambiguous areas (blue squares) in the top and bottom rows caused by the same pathological variability and the partial volume effect were annotated differently. The images are sliced and zoomed from 3D images shown in the right-bottom corners for visibility.}
  \label{fig:Ambiguous_annot}
\end{figure}

\begin{figure*}[t!]
\centering
  \subfloat{ \includegraphics[width=175mm]{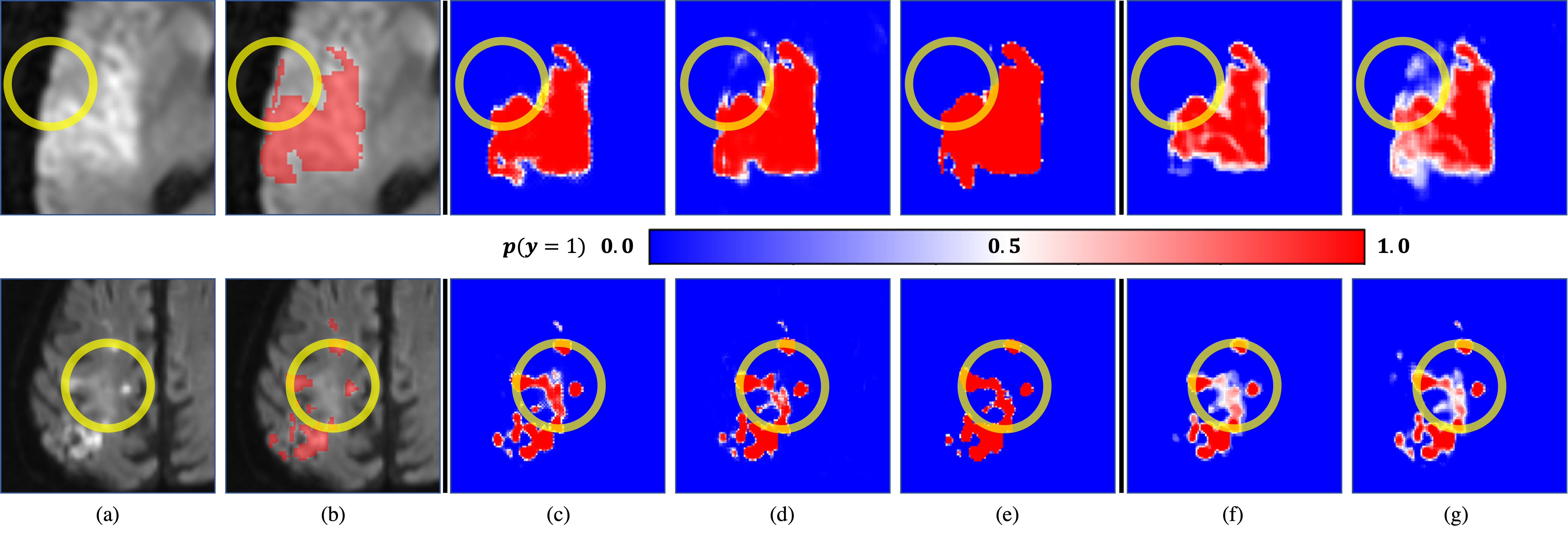} } 
\vspace*{-4mm}
\caption{The comparisons of the estimated segmentation probability maps. (a) input images, (b) human annotations, (c) ResUNet, (d) UNETR, (e) UNETR ensembles with different data subsets, (g) UNETR ensembles with the varying Tversky loss (ours), and (h) Hypernetwork ensembles with varying Tversky loss (ours). The ambiguous areas caused by image artifacts (yellow circle, top row) and different pathology (yellow circle, bottom row) were both estimated to have 0.5 probability by the proposed methods(f-g).}
\label{fig:ProbEstimates}
\end{figure*}

\section{Introduction}
\label{sec:intro}

Deep learning (DL) based methods have been the prime choices for image segmentation tasks in the past decade~\cite{lecun2015deep,minaee2021image,taghanaki2021deep,ronneberger2015u}.
Despite their superior performance, DL networks are notoriously overconfident about their predictions that the estimated probabilities of the predicted segmentation labels are polarized close to zero or one~\cite{ronneberger2015u,lee2016stochastic,kohl2018probabilistic,monteiro2020stochastic}. 

In many applications, especially in medicine, the polarized segmentation probability is often not desirable because there are many factors that cause the inherent ambiguity in annotations even for experienced human experts, e.g., the partial-volume effect, the continuous lesion infarction process, or annotator fatigue~\cite{fischl2002whole,kohl2018probabilistic,liao2020iteratively,thomalla2011dwi,soret2007partial,billot2020partial,baldwin2004sleep,sagonas2013semi}. 
Fig.~\ref{fig:Ambiguous_annot} (b) shows the examples of the human annotations of ambiguous stroke lesions in two diffusion weighted images (DWIs, Fig.~\ref{fig:Ambiguous_annot} (a))~\cite{giese2017design}.
Ambiguous lesion areas (blue squares) caused by the lesion diffusion in the image at the top row were annotated by a human expert while not annotated in another image at the bottom row.

Many approaches were proposed to overcome this challenge by leveraging the segmentation uncertainty and multiple annotations per image. 
The network ensemble approaches that ensemble the outputs of multiple DL networks trained by different configurations~\cite{hansen1990neural,krogh1995neural,zhou2021ensemble,ganaie2021ensemble,sundaresan2021triplanar,li2018fully,chen2021one,li2021white,herron2020ensembles,garipov2018loss,zhang2019fully} and implicit ensemble approaches~\cite{srivastava2014dropout,huang2016deep,ganaie2021ensemble} were proposed to improve the generalization of the DL networks. However, their uncertainty maps did not fully represent the underlying structures of objects. 
Another popular choice is the M-heads method that estimates multiple label candidates~\cite{lakshminarayanan2017simple,rupprecht2017learning,ji2021learning}.
The stochastic networks are also popular choices to model the uncertainty and generate multiple label candidates~\cite{sagonas2013semi,kohl2019hierarchical,baumgartner2019phiseg,kendall2015bayesian,monteiro2020stochastic}. However, the methods often rely on multiple annotations per image and do not directly offer a segmentation probability map that can provide an intuitive control to users to adjust the estimated segmentation label maps.

In this paper, we aim to improve the segmentation probability estimation for medical image segmentation in a common real-world scenario that we only have one ambiguous annotation per each image without sacrificing the performance of the DL networks on the binary segmentation.
We choose to model segmentation probability maps directly rather than modeling the uncertainty to provide intuitive control on different segmentation results to end-users. 

Our hypothesis is that we can estimate the underlying segmentation probability map by marginalizing the segmentation label maps estimated by networks that are encouraged to under-/over-segment while not penalizing the balanced segmentation. 
We first propose the network ensemble approach that leverages the varying Tversky loss~\cite{salehi2017tversky}. 
The Tversky loss was first suggested for medical image segmentation to tackle the imbalanced segmentation problem~\cite{salehi2017tversky}.
We leverage the flexibility of the Tversky loss that can encourage the network to under-/over-segment without penalizing the balanced segmentation by varying its hyperparameters. 
To overcome the high computational requirement of the proposed network ensemble approach, we also propose a unified approach leveraging a hypernetwork architecture~\cite{ha2016hypernetworks,hoopes2021hypermorph} that learns different configurations of the primary network with respect to the hyperparameter of the varying Tversky loss. 
We show the feasibility and strength of our proposed methods and compare them to the state-of-the-arts on 3D medical images in application to the acute stroke lesion segmentation with clinical-grade DWIs collected from multiple hospitals~\cite{giese2017design}.


Our contributions can be largely summarized as follows: 
1) We improve the segmentation probability map (shown in Fig.~\ref{fig:ProbEstimates}) estimation to reflect underlying structures from a single ambiguous annotation per image without sacrificing the binary segmentation performance.
2) It is the first attempt in our knowledge to leverage the combinations of the network and hypernetwork ensemble approaches with varying Tversky loss to improve the segmentation probability estimation.
3) The propose methods offer an intuitive control on the binary segmentation results by simply choosing different probability thresholds. 


\subsection{Related works}
In this subsection, we will summarize the previous works that our proposed methods are built upon. 

\noindent \textbf{Medical Image Segmentation }There are a plethora of papers on medical image segmentation as it is a fundamental task in medicine to delineate anatomical structures~\cite{fischl2012freesurfer,pham2000current,shen2017deep}.
Since the UNet~\cite{ronneberger2015u}, the DL-based approaches have become the standard in medical image segmentation and shown successful results~\cite{shen2017deep,kamnitsas2016deepmedic,oktay2018attention}. 
Because the data imbalance and the label ambiguity, many approaches were suggested to manipulate loss functions to achieve better performance, e.g., the Tversky loss~\cite{salehi2017tversky,lin2017focal,bertels2019optimization,abraham2019novel}. 
Recently, the multi-head transformer-based approach showed the-state-of-the-art performance with 3D brain images~\cite{hatamizadeh2021unetr}. 

\noindent \textbf{Hypernetworks } The hypernetwork architecture was suggested in~\cite{ha2016hypernetworks} to mitigate the computational burden of DL networks by estimating the weights of a primary network. 
It was applied to many applications including image recognition, neural architectural search and neural representations~\cite{ha2016hypernetworks,zhang2018graph,jia2016dynamic,klocek2019hypernetwork,littwin2019deep,nirkin2021hyperseg}.
The hypernetwork was used in medical applications to alleviate the computational burden of the hyperparameter search for image registration and reconstruction~\cite{hoopes2021hypermorph,wang2021hyperrecon}. 
It was also applied to real-time semantic segmentation and medical image segmentation and showed successful results~\cite{nirkin2021hyperseg,ma2021hyper}.

\noindent \textbf{Network Ensembles } The network ensemble is a popular approach to improve the generalization of DL networks~\cite{hansen1990neural,krogh1995neural,zhou2021ensemble,ganaie2021ensemble}.
Typically, the network ensemble approaches include: 1) the $k$-fold cross-validation strategy that trains multiple networks with different subsets of training data and random initialization of the networks~\cite{krogh1995neural,sundaresan2021triplanar,li2018fully}, 2) different network/training configurations~\cite{chen2021one,li2021white,herron2020ensembles,garipov2018loss,zhang2019fully,zhou2021ensemble}, and 3) \textit{implicit} ensemble with dropout-like schemes~\cite{srivastava2014dropout,huang2016deep,ganaie2021ensemble}.
The reasoning behind employing the network ensemble is that DL networks are easy to converge to local minima because of the large number of parameters. 
The previous approaches rely on the stochastic characteristics of initialization and data selection in different network training configurations while an individual network is trained on the same objective to achieve the best performance on a given task that may not be optimal for the probability estimation. 

\begin{figure*}[t]
\centering
  \subfloat{ \includegraphics[width=175mm]{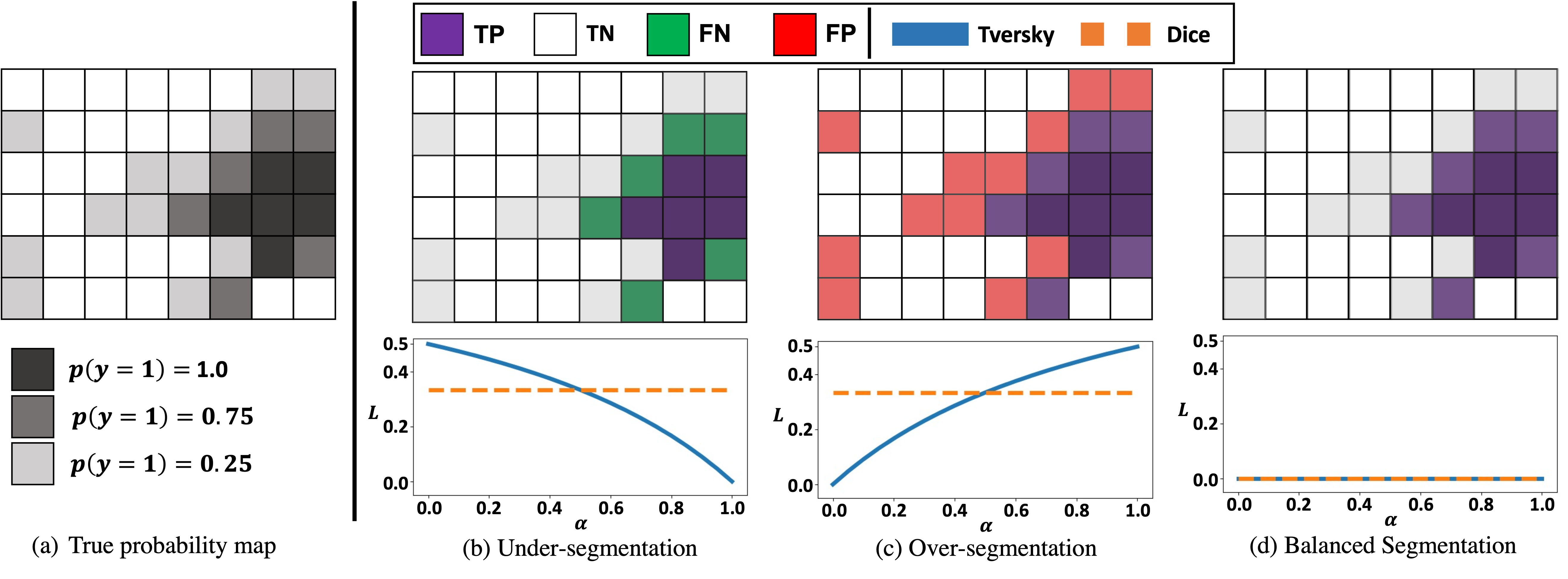} } 
\vspace*{-4mm}
\caption{(Top) The illustrations of under- and over-segmentation with the hypothetical \textit{true} probability map with true positive (TP, purple, true negative (TN, white), false negative (FN, green), and false positive (FP, red) labels. 
(Bottom) The corresponding Tversky loss $L$ with varying hyperparameter $\alpha$. The undersegmentation is less penalized with higher $\alpha$ while the oversegmentation is less penalized with lower $\alpha$. The balanced segmentation is always encouraged regardless the choice of $\alpha$.}
\label{fig:UnderOverSeg}
\end{figure*}

\section{Varying Tversky Loss}
\label{sec:Tversky}

Let us assume that we have an underlying \textit{true} segmentation probability map as illustrated in Fig.~\ref{fig:UnderOverSeg} (a) from many human experts for binary segmentation.
With the \textit{true} probability map, the binary segmentation can be done by simple thresholding.
It can be simply formulated as follows,
\begin{equation}
    \bm{P} = \delta( p( \bm{y} ), \tau ),
\label{eq:prob_thres}
\end{equation}
where $\bm{P}{\in}\{0,1\}^{W{\times}H{\times}D}$ is a predicted segmentation label and $p(\bm{y})$ is the \textit{true} probability map of the potential segmentation labels $\bm{y}{\in}\{0,1\}^{W{\times}H{\times}D}$ for a 3D image with the height $H$, width $W$, and depth $D$. $\delta$ is the thresholding function with a threshold $\tau$. 
Low $\tau$ will lead to undersegmentation as shown in the top row of Fig.~\ref{fig:UnderOverSeg} (b), and high $\tau$ to oversegmentation as shown in the top row of Fig.~\ref{fig:UnderOverSeg} (c). 
Our intuition behind our approach is simple. Since we do not have the \textit{true} probability map and DL networks tend to estimate the highly polarized segmentation probability, we \textit{flip} the problem in Eq.~\ref{eq:prob_thres} to estimate the segmentation probability map by marginalizing predicted segmentation probability maps from multiple networks encouraged to under-/over-segment while not penalizing balanced segmentation (Fig.~\ref{fig:UnderOverSeg} (d)). 
It can be formally stated as follows, 
\begin{equation}
    p( \bm{y} ) = \int p(\bm{y};{\bm{\theta}|\bm{h}}) p( \bm{h} ) d\bm{h},
\end{equation}
where $\bm{h}$ is a hyperparameter to enforce a network to estimate under-/over-segmentation and $p( \bm{y}; {\bm{\theta}|\bm{h}})$ is the polarized probability maps estimated by a network $\bm{\theta}$ conditioned on $\bm{h}$.
For simplicity, we assume that $\bm{\theta}$ is determined by $\bm{h}$ without stochasticity. 
One necessary property of $\bm{\theta}|\bm{h}$ is that it should not penalize the balanced segmentation to guarantee training stability.
Otherwise, it may induce undesired extra ambiguity to the segmentation probability estimation.

We leverage the Tversky loss function that scales the effects of false positives (FP) and false negatives (FN) in total loss~\cite{salehi2017tversky}.
The Tversky index $S_{Tv}$ is formulated as follows,
\begin{equation}
    S_{Tv}( \bm{P}, \bm{G} ;\alpha, \beta) = \frac{ \bm{P} \cap \bm{G} }{ \bm{P} \cap \bm{G} + \alpha \bm{P} \setminus \bm{G} + \beta \bm{G} \setminus \bm{P}},
\label{eq:Tversky}
\end{equation}
where $\bm{P}$ and $\bm{G}$ are a predicted label and a ground truth, respectively. It has two hyperparameters, $\alpha$ and $\beta$, controlling the effects of FP, $\bm{P} {\setminus} \bm{G}$, and FN, $\bm{G} {\setminus} \bm{P}$, respectively~\cite{salehi2017tversky,tversky1977features}. 
When $\alpha{=}0.5$ and $\beta{=}0.5$, the $S_{Tv}$ becomes the Dice index $S_{Dice}$ that the effects of FP and FN are equally considered~\cite{dice1945measures,salehi2017tversky}. 
$S_{Tv}$ is used as a loss function for image segmentation similar to the soft Dice loss~\cite{salehi2017tversky},
\begin{equation}
    L_{Tv}( \bm{P}, \bm{G}; \alpha, \beta) = 1 - S_{Tv}( \bm{P}, \bm{G}; \alpha, \beta).
\label{eq:TvLoss}
\end{equation}

We focus on the flexibility of the Tversky index $S_{Tv}( P, G;\alpha, \beta)$ that can enforce a network to learn to under-/over-segment rather than finding the optimal values of $\alpha$ and $\beta$.
We constrain $\alpha$ and $\beta$ to be $0{\leq}\alpha {\leq}1$, $\beta{=} 1.0{-}\alpha$. This constraint makes $S_{Tv}{\in}[0,1]$ commensurate with the range of the Dice coefficient. In this way, $\alpha$ and $\beta$ become the balancing weights between FP and FN. 

In the bottom row of Fig.~\ref{fig:UnderOverSeg} (b), it shows that the effect of the undersegmentation is less reflected in the Tversky loss function with higher $\alpha$ (i.e., lower $\beta$). In an extreme case $\alpha=1.0$, the Tversky loss becomes 0.0 that undersegmentation is considered to be the same as perfect segmentation. 
On the other hand, the effect of the oversegmentation is less reflected in the Tversky loss function with lower $\alpha$ as shown in Fig.~\ref{fig:UnderOverSeg} (c). 
In both cases, the soft Dice loss stays the same because the effects of FP and FN are equally considered. 
The Tversky loss function returns 0.0 regardless of the choice of $\alpha$ for the correct and balanced segmentation as shown in Fig.~\ref{fig:UnderOverSeg} (d). This is highly desirable for our purpose for the segmentation probability estimation since we do not want to penalize the balanced segmentation.  

We enforce a network to learn to under-/over-segment by varying $\alpha$ and $\beta$: i.e., $\alpha > \beta$ to undersegment and $\alpha < \beta$ to oversegment.
Note that the same purpose can be achieved by manipulating weights of the log losses of positive and negative labels in the binary cross-entropy function~\cite{cox1958regression}.
We chose the Tversky loss because it was normalized nicely and shown to perform well for medical image segmentation.



\section{Ensemble with Varying Tversky Loss}
\label{sec:Ensemble}

One straightforward approach to estimate the segmentation probability maps with the varying Tversky loss is the network ensembles with multiple networks trained with varying Tversky loss hyperparameters, $\alpha$ and $\beta$~\cite{hansen1990neural}.

For a standard DL approach, a segmentation probability estimation function $f( \bm{I}| \bm{\theta}) = p( \bm{y}; \bm{\theta} )$ estimates the segmentation probability map $p(\bm{y}; \bm{\theta})$ conditioned on the network $\bm{\theta}$ from an image $\bm{I}$. 
The ensemble approach with multiple networks $\bm{\Theta} = \{ \bm{\theta}_1, ..., \bm{\theta}_{N_{e}} \}$ is formulated as follows,
\begin{align}
    f_{ens} (\bm{I} ) &= \int f(\bm{I} | \bm{\theta} ) p( \bm{\theta} )  d \bm{\theta} \\
    &= \frac{1}{N_{e}} \sum_{i=1}^{N_{e}} f( \bm{I} | \bm{\theta}_i ), \ if \ p(\bm{\theta}_i)=\frac{1}{N_{e}}, \forall i, 
\end{align}
where $N_{e}$ is the number of the individual networks.

Let $\bm{H} = \{ \bm{h}_1, ..., \bm{h}_{N_{e}} \}$ and $\bm{\Theta} = \{ \bm{\theta}_1, ..., \bm{\theta}_{N_{e}} \}$ be the sets of varying Tversky loss hyperparameters, $\bm{h}_i = ( \alpha_i, \beta_i )$, and the DL networks determined by the corresponding $\bm{h}_i$, respectively.
The hyperparameter $\bm{h}$ is in the 1-simplex because of the constraint, $0{\leq}\alpha_i{\leq}1$, $\beta_i{=}1 {-} \alpha_i$. 
The network $\bm{\theta}_i$ is trained to penalize less on undersegmentation with $\alpha_i{>}\beta_i$, and less on oversegmentation with $\beta_i{>}\alpha_i$. 
The segmentation probability estimation of each network is now conditioned on the hyperparamters $\bm{h}_i$ since an individual network $\bm{\theta}_i$ is trained with the Tversky loss $L_{Tv}(\cdot;\bm{h}_i)$,
\begin{equation}
    f( \bm{I}; \bm{\theta}_i | \bm{h}_i ) = p( \bm{y}; \bm{\theta}_i | \bm{h}_i ).
\label{eq:segprob_tv}
\end{equation}

The ensemble function $f_{ens}$ is defined as follows, 
\begin{equation}
    f_{ens}( \bm{I} ) = \frac{1}{N_e} \sum_{i=1}^{N_e} f( \bm{I}; \bm{\theta}_i | \bm{h}_i ).
\label{eq:ensemble_tv}
\end{equation}
It is the discrete version of $\bm{p}_{ens}(\bm{y})=\int  \bm{p}( \bm{y}; \bm{\theta}|\bm{h} ) p(\bm{h}) d\bm{h}$ over $\bm{h}$ when $\bm{h}$ is uniformly distributed, i.e., $p(\bm{h}_i)=1/N_{e}, \ \forall i=1,...,N_{e}$.


Our approach relies on varying segmentation results estimated by DL networks conditioned on the varying Tversky loss without manipulating human annotations or any fixed priors on the segmentation labels to leverage the flexibility of DL networks on segmentation.
One drawback of our ensemble method is that it requires high computational resources to train multiple networks. Typically, the number of individual networks for the ensemble is limited (e.g., three to five) which is not optimal for the probability estimation.

\begin{figure*}[tb!]
\centering
  \subfloat{ \includegraphics[width=140mm]{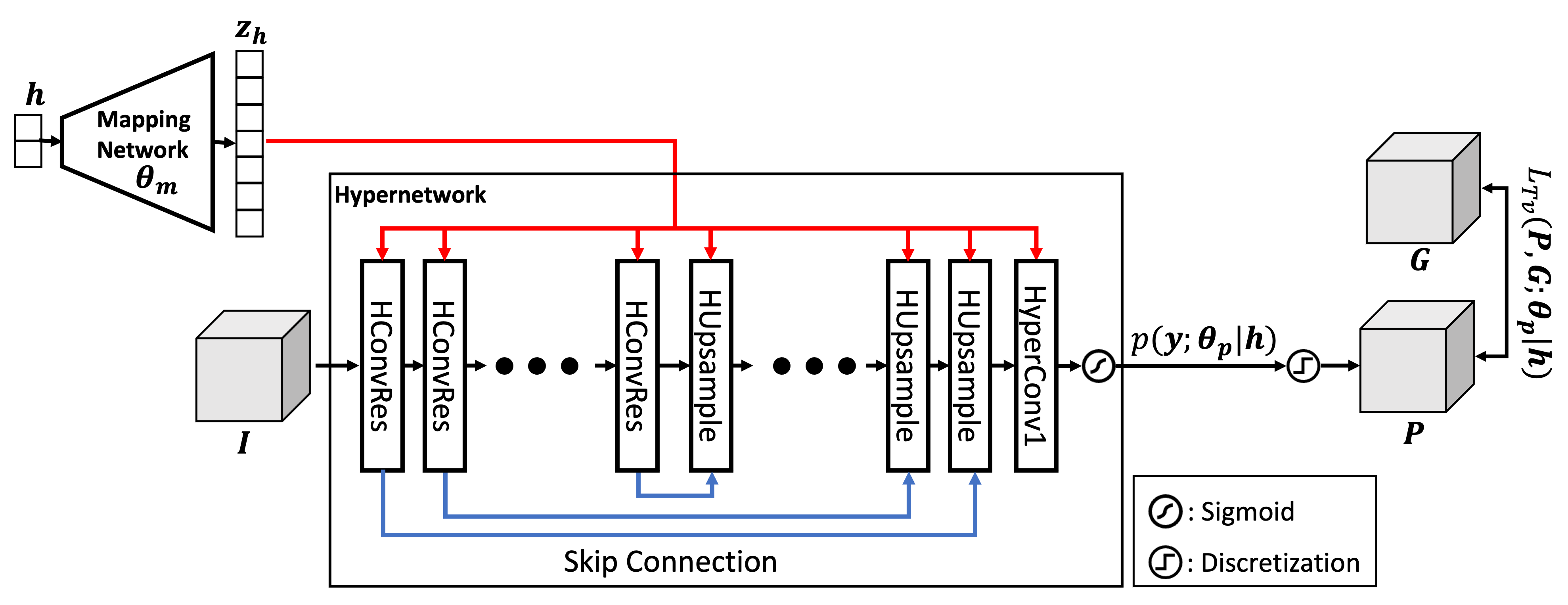} } 
\vspace{-3mm}
\caption{Hypernetwork architecture. The hypervector $\bm{z_h}$ mapped by the mapping network $\bm{\theta_m}$ from a hyperparameter $\bm{h}$ is used to generate the parameters of the primary ResUNet network $\bm{\theta_p}$ in the hyperconvolution residual (HConvRes) and upsample (HUpsample) blocks. The primary network takes an image $\bm{I}$ and estimate the segmentation label map $\bm{P}$. The final feature output of $\bm{\theta_p}$ is transformed to the segmentation probability map $p(\bm{y};\bm{\theta_p}|\bm{h})$ by the sigmoid and discretized to $\bm{P}$. During training, the loss is calculated by the Tversky loss with the hyperparameter $\bm{h}$ between $\bm{P}$ and a human annotation $\bm{G}$.}
\label{fig:Hypernet}
\end{figure*}

\begin{figure*}[tb!]
\centering
  \subfloat{ \includegraphics[width=170mm]{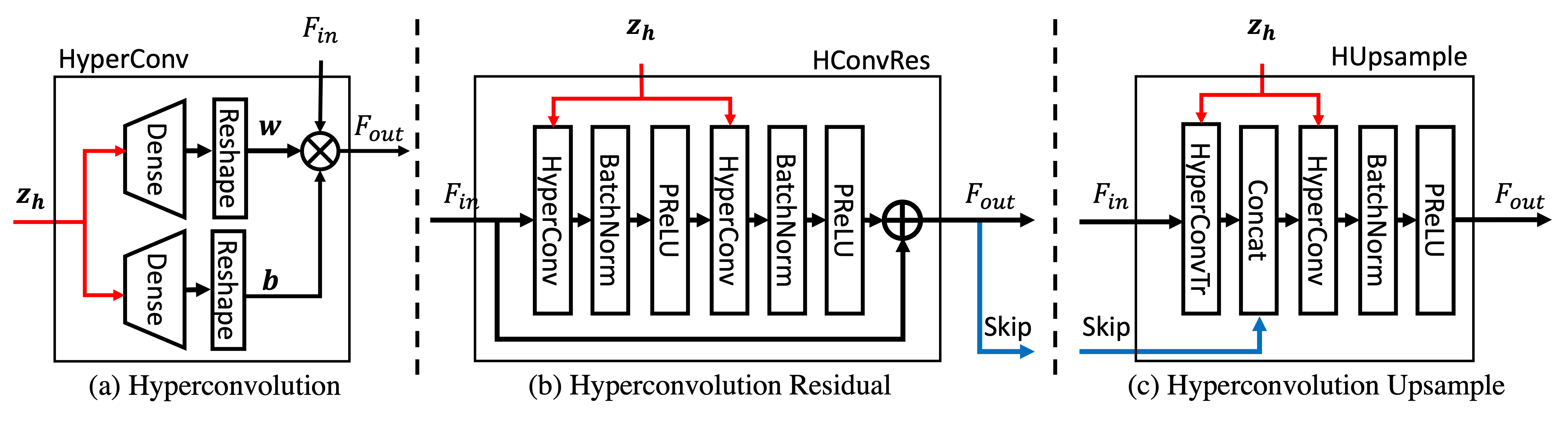} }
\vspace*{-3mm}
\caption{Hyperblocks. (a) The hyperconvolution (HyperConv) block estimates the kernel weights $\bm{w}$ and biases $\bm{b}$ of the convolution block of the primary network via the separate dense layers. (b) The HyperConv residual unit (HConvRes) for the encoding part of the primary network is the sequence of the HyperConv, batch normalization, and PReLU. (c) The hyperupsample (HUpsample) for the decoder consists of the transposed HyperConv (HyperConvTr), concatenation, HyperConv, batch normalization, and PReLU.}
\label{fig:Hyperblocks}
\end{figure*}

\section{Hypernetwork Ensemble}
\label{sec:Hyper}


\begin{figure*}[t]
\centering
  \subfloat{ \includegraphics[width=175mm]{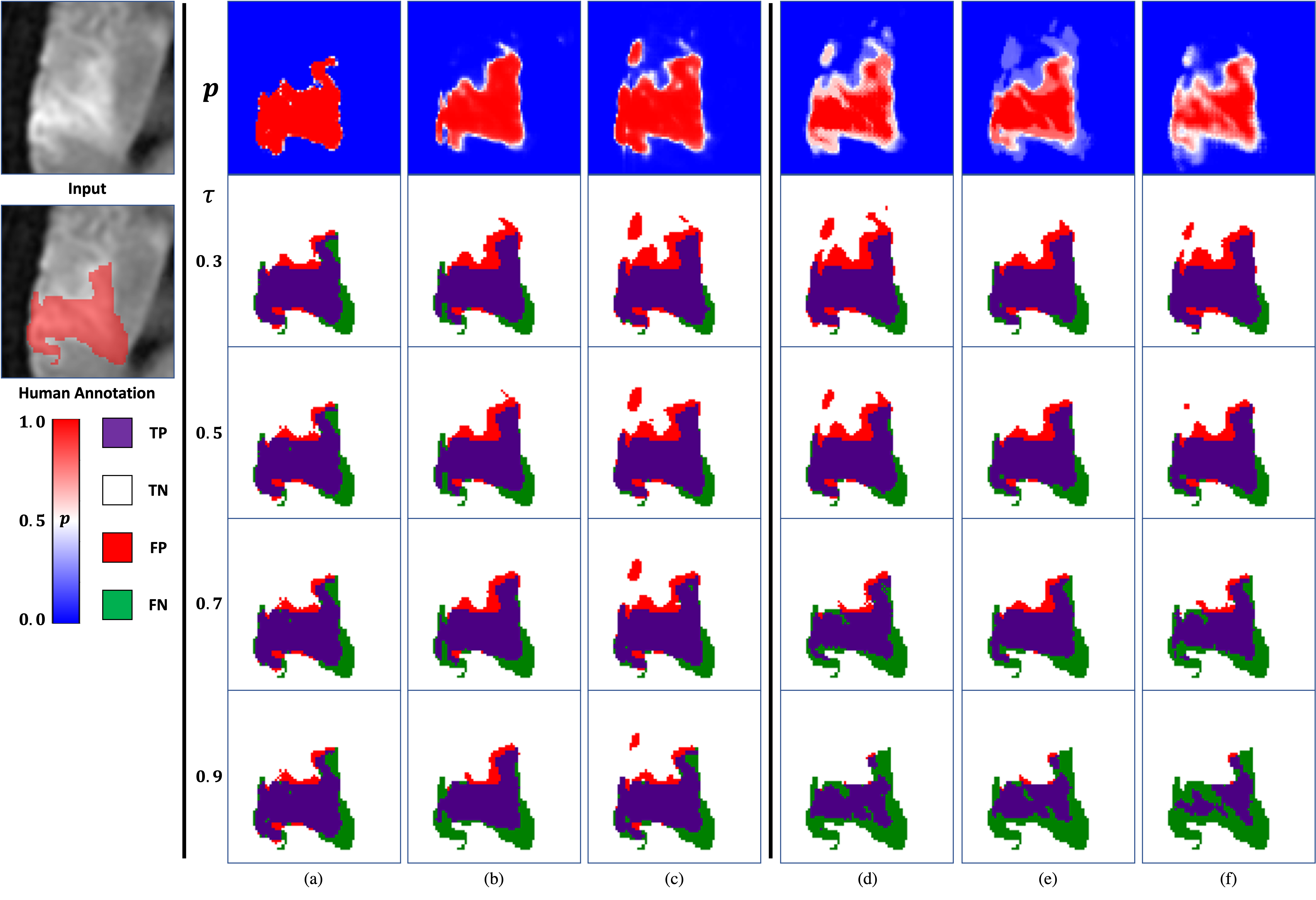} } 
\vspace*{-4mm}
\caption{Binary segmentation results with varying probability thresholds $\tau$. The probability maps estimated by (a) UNETR, (b) UNETR with dropout, (c) UNETR ensemble with random data subsets, (d) our ResUNet ensemble with varying Tversky loss (V-Tv), (e) UNETR ensemble with V-Tv, and (f) hypernetwork ensemble methods. Our methods (e-f) captured the underlying lesion structure.}
\label{fig:ThresHardSeg}
\end{figure*}

\subsection{Hypernetwork Architecture}
\label{ssec:hyper_resunet}

The hypernetwork architecture in our framework is adapted from the hypernetwork in~\cite{hoopes2021hypermorph} and the residual UNet in~\cite{kerfoot2018left}.
Fig.~\ref{fig:Hypernet} illustrates the overall architecture of the proposed hypernetwork. 
The mapping network $\bm{\theta_m}$ generates the hypervector $\bm{z_h}$ from the hyperparameter $\bm{h}{=}(\alpha, \beta)$. The mapping network consists of dense layers and rectifying linear units (ReLU)~\cite{karras2019style,hoopes2021hypermorph}.

The hypernetwork consists of hyperconvolution residual (HConvRes) and hyperconvolution upsample (HUpsample) units with skip connections that constitute the ResUNet architecture with the hyperconvolution (HyperConv) blocks substituting the standard convolution blocks~\cite{kerfoot2018left,ronneberger2015u,hoopes2021hypermorph,lecun1995convolutional}.
The HyperConv block is illustrated in Fig.~\ref{fig:Hyperblocks} (a). 
There are two single dense layers that take the mapped hypervector $\bm{z_h}$ as an input. 
Those dense layers estimate the kernel weight $\bm{w}$ and bias $\bm{b}$ for the convolution.
The input feature $F_{in}$ is convolved with the generated $\bm{w}$ and $\bm{b}$. 
The parameters of the dense layers and the generated paramters of the convolution blocks constitute the hypernetwork $\bm{\theta_h}$ and the primary network $\bm{\theta_p}$, respectively. 
The transposed HyperConv (HyperConvTr) is similar to the HyperConv block with the transposed shape of $\bm{w}$ and the transposed convolution function instead of the convolution function.
The HConvRes and HUpsample units are illustrated in Fig.~\ref{fig:Hyperblocks} (b) and (c), respectively. The residual and upsample structures with the skip connection are adapted from~\cite{kerfoot2018left} with the parametric rectifying linear units (PReLUs) and the batch normalization~\cite{zhang2018road,he2015delving,ronneberger2015u}.
The hypernetwork parameters $\bm{\theta_h}$ are trainable parameters while the parameters of the primary segmentation network $\bm{\theta_p}$ are conditioned on the hyperparameters $\bm{h}$ and generated by $\bm{\theta_h}$.
After the encoding and decoding steps, the features are convolved with the 1x1x1 HyperConv block followed by the sigmoid function to generate a segmentation probability map $p(\bm{y};\bm{\theta_p}|\bm{h})$. 

At the training stage, the final segmentation label map $\bm{P}$ is obtained by discretizing the the segmentation probability map by simple thresholding for binary segmentation.
We used the Tversky loss function $L_{Tv} (\bm{P}, \bm{G}; \bm{h})$ in Eq.~\ref{eq:TvLoss} with randomly sampled $\bm{h}{=}( \alpha, \beta)$ where $\alpha{\sim}\textrm{U}(\epsilon, 1-\epsilon)$, $\beta = 1-\alpha$ for each minibatch as a training loss function. $\epsilon$ is the small value to guarantee computational stability. We set $\epsilon{=}0.05$ for experiments.

\begin{table*}[t]
\centering
\begin{tabular}{llcccccc}
\toprule
\multicolumn{2}{l}{}                                     & Dice          & Bal. Acc.     & Precision     & Recall        & ROC AUC   & \makecell{Training Time \\(GPU Hours)}     \\ 
\toprule
\multirow{3}{*}{Baselines}              & ResUNet (Dice) & 0.803          & 0.873          & 0.869          & 0.747          & 0.822    &    \textbf{7}  \\
                                        & ResUNet        & 0.807          & 0.876          & 0.870          & 0.752          & 0.856    &    8  \\
                                        & UNETR          & 0.797          & 0.865          & 0.878          & 0.730          & 0.851    &    38  \\ \hline
\multirow{2}{*}{Dropout}                & ResUNet        & 0.803          & 0.882          & 0.848          & 0.763          & 0.819    &    8  \\
                                        & UNETR          & 0.807          & 0.871          & 0.884          & 0.743          & 0.857    &   57   \\ \hline
\multirow{2}{*}{Ensemble}               & ResUNet        & 0.806          & 0.874          & 0.873          & 0.747          & 0.823    &    37  \\
                                        & UNETR          & 0.795          & 0.855          & \textbf{0.904} & 0.709          & 0.854    &    190  \\ \hline
\multirow{2}{*}{Ensemble w. VTv (Ours)} & ResUNet        & \textbf{0.820} & 0.888          & 0.868          & 0.776          & \textbf{0.884}& 53 \\
                                        & UNETR          & 0.799          & 0.872          & 0.862          & 0.745          & 0.863    & 258     \\ \hline
Hypernet w. VTv (Ours)                  & HyperUNet      & 0.811        & \textbf{0.891} & 0.842          & \textbf{0.783} & 0.869  & 13       
\\
\bottomrule
\end{tabular}
\vspace*{-2mm}
\caption{Binary segmentation results with the fixed probability threshold (0.5) and the training time (GPU hours). The best performance metrics are bold-faced.}
\label{tab:hard_res}
\end{table*}

\begin{figure}[htb!]
\centering
  \subfloat[Dice scores]{ \includegraphics[width=80mm]{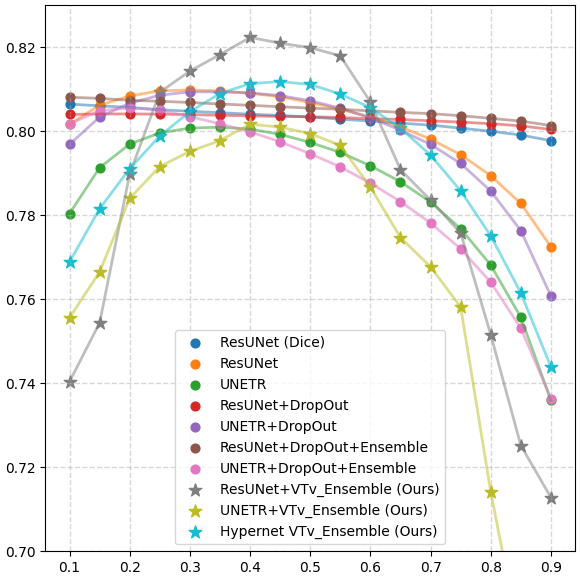} } \\
  \subfloat[ROC curves]{ \includegraphics[width=80mm]{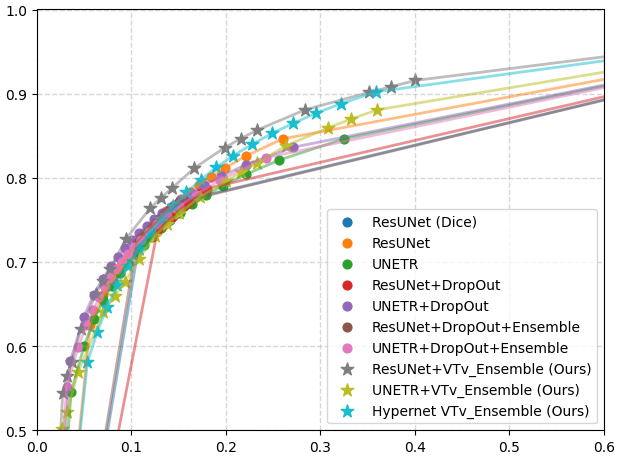} } 
\vspace*{-3mm}
\caption{(a) Dice and (b) receiver operating curves (ROC) with varying probability thresholds. The proposed ResUNet ensemble (gray stars) and the hypernetwork (cyan stars) ensemble with the varying Tversky loss showed the best performance.}
\label{fig:ProbThres_Dice}
\end{figure}

\subsection{Hypernetwork Ensemble}
\label{ssec:HyperEnsemble}

The ensemble process with the hypernetwork is similar to the network ensemble strategy with the varying Tversky loss described in Section~\ref{sec:Ensemble}.
The advantage of the hypernetwork ensemble over the network ensemble strategy is that we can sample hyperparameters $\bm{h}$ with any intervals with a single hypernetwork because the the hypernetwork is trained on continuously sampled $\bm{h}$ at the cost of the increased inference time proportional to the number of $\bm{h}$ for the ensemble, $O(N_{e})$.
The set of multiple DL networks $\bm{\Theta}$ in Eq.~\ref{eq:ensemble_tv} is now replaced with the estimated parameters of the primary segmentation network $\bm{\theta_p}$ determined by $\bm{h}_i$,
\begin{equation}
    f_{hyper}( \bm{I} ) = \frac{1}{N_{e}} \sum_{i=1}^{N_{e}} f( \bm{I}; \bm{\theta_p} | \bm{h}_i ). 
\label{eq:hyper_ensemble_tv}
\end{equation}
Finally, the segmentation probability map $p(\bm{y})$,  estimated by $f_{hyper}(\bm{I}) = p(\bm{y})$, is binarized by thresholding to generate a binary segmentation label map. 

\section{Experiments}
\label{sec:Experiments}

\noindent \textbf{Dataset } We applied our proposed methods to the acute ischemic stroke lesion segmentation problem with clinical-grade 3D diffusion weighted images (DWIs) from the MRI-GENIE study~\cite{giese2017design}. 
The informed and written consent forms were obtained from all patients or their legal representatives~\cite{giese2017design,meschia2013stroke}. Each hospital received approval of their internal review board~\cite{giese2017design}. 
Five hundreds and fifty images with human expert annotations were used in our experiments. We divide the data set to 412 ($\sim$75\%) training data and 138 ($\sim$25\%) validation data.
All images were center-aligned and resampled to 1.0x1.0x6.0$mm^3$ (256x256x32$vox$). The intensity range of each image was scaled to [0,1]. 

\noindent \textbf{Configurations } We compared the Residual UNet (ResUNet)~\cite{ronneberger2015u,kerfoot2018left} and the multi-head transformer-based method (UNETR)~\cite{hatamizadeh2021unetr} that showed the state-of-the-art performance for 3D medical images to our methods.
We trained the ResUNet and UNETR in three configurations: 1) the \textit{vanilla} setting, 2) dropout with the rate of 0.1, and 3) the network ensembles with 5-fold training subsets. 
We trained all comparison methods with the Dice cross-entropy (CE) loss~\cite{taghanaki2019combo}. 
One vanilla ResUNet was trained with the soft Dice loss as a baseline~\cite{dice1945measures}.
Our network ensemble approach with the varying Tversky (V-Tv) loss used the separate ensembles of five individual ResUNet and UNETR networks with the same training set to eliminate the effect of random data.
We trained individual networks with the Tversky loss hyperparamters $\alpha=\{ 0.1, 0.3, 0.5, 0.7, 0.9\}$. 

\noindent \textbf{Implementation Details } For all configurations, images and annotation labels were randomly cropped to 192x192x16 patches. 
At the inference stage, the segmentation probability maps and label maps were predicted by the sliding windows technique with 80\% overlap~\cite{keogh2001online}. 
We optimized all configurations except the hypernetwork approach with the Adam optimizer with the learning rate $1e^{-4}$ and the weight decay $1e^{-3}$~\cite{kingma2014adam}.
The hypernetwork was trained with the learning rate $1e^{-5}$ and the weight decay $1e^{-3}$ for training stability. 
We implemented our hypernetwork architecture with the MONAI 0.7.0 and PyTorch 1.9 frameworks\footnote{https://monai.io/, https://pytorch.org/}~\cite{monai2020,NEURIPS2019_9015}. We used one NVIDIA Titan Xp with 12 GB GPU memory per network for training and inference. 


\subsection{Segmentation Probability Estimation}
\label{ssec:res_seg_prob}

Fig.~\ref{fig:ProbEstimates} and the top row of Fig.~\ref{fig:ThresHardSeg} show the examples of the estimated segmentation probability maps. 
In Fig.~\ref{fig:ProbEstimates}, the segmentation probability maps of a validation image (Fig.~\ref{fig:ProbEstimates} (a)) with a human annotation (Fig.~\ref{fig:ProbEstimates} (b)) estimated by the vanilla ResUNet, UNETR with the dropout, UNETR ensembles with Dice CE loss with data subsets, and our proposed UNETR ensemble with the V-Tv loss and hypernetwork ensemble approach (Fig.~\ref{fig:ProbEstimates} (c-g), respectively) are presented. 
The segmentation probability maps estimated by the other methods were highly polarized to 0.0 and 1.0 which did not reflect the ambiguity.
At the bottom of Figure~\ref{fig:ProbEstimates}, an ambiguous area (yellow circle) between the stroke lesions was segmented as stroke lesions with high confidence while the area is the white matter hyperintensity (WMH) which is different from stroke lesions. Our approaches estimated the probability close to 0.5 for the area that indicated the ambiguity between WMH and lesions. The example in the top row shows the strength of our method for the ambiguity between lesion and possible image artifact (yellow circle).
The qualitative example shown in Fig.~\ref{fig:ThresHardSeg} shows the strength of our approaches that we can control the binary segmentation results with different probability thresholds. 
Our estimated probability maps captured the highly probable area of lesions as shown in the input image on the left column with the high probability threshold (the bottom row Fig.~\ref{fig:ThresHardSeg} (d-f)).
Different thresholds affected the results from the other methods insignificantly(Fig.~\ref{fig:ThresHardSeg} (a-c)). 

Fig.~\ref{fig:ProbThres_Dice} (a) shows the curves of aggregated Dice scores over different probability thresholds.
Because there is the inherent ambiguity in our human annotations, higher Dice scores can be relatively less important in our evaluation than the other segmentation tasks.
However, our proposed ResUNet ensmeble with the V-Tv loss and hypernetwork methods (marked by stars) outperformed the other methods (marked by circles).
The important characteristic of the proposed methods shown in the Dice curves is that the Dice score changes smoothly over probability thresholds. 
This aligns well with the intuition that the higher probability thresholds would lead to undersegmentation and the lower to oversegmentation when the \textit{true} probability map is given. 
As shown in Fig.~\ref{fig:ThresHardSeg}, this property gives the control to users that they can decide the level of under-/over-segmentation. 

The receiver operating characteristic (ROC) curves in Fig.~\ref{fig:ProbThres_Dice} (b) shows how the false positive rates (FPR) and the true positive rates (TPR) are distributed with the same 0.05 probability threshold interval for the proposed methods (marked by stars) and the other methods (marked by circles). More densely clustered FPR-TPR points with varying thresholds of the other approaches indicated the more polarized segmentation probability estimation. 
The proposed ResNet ensemble with the V-Tv loss achieved the best area under curve of the ROC (ROC-AUC) followed by the hypernetwork ensemble as shown in Table~\ref{tab:hard_res}. 
Because of the imbalanced ratios of positive and negative labels (i.e., smaller lesion volumes compared to the whole brain), the TPR and FPR did not go smoothly to 1.0 for all methods. 

\subsection{Binary Segmentation Results}
\label{ssec:res_binary_seg}

We compared the binary segmentation results by setting the segmentation probability threshold to 0.5. 
Table~\ref{tab:hard_res} summarizes the evaluation metrics of the proposed and the other methods. 
The proposed ResUNet ensemble with the V-TV loss achieved the best performances on the aggregated Dice score (0.82) and ROC-AUC (0.884).
The hypernetwork ensemble achieved the best performance on the balanced accuracy (0.891) and recall (0.783) and the second best on the ROC-AUC (0.869). 
The UNETR ensembles with the random data subsets achieved the best precision (0.904). Its relatively low recall (0.709) may indicate that it enforced undersegmentation. 
The binary segmentation results showed that our methods did not sacrifice the performance on the binary segmentation in the full-automated scenario. 
They marginally improved the results compared to the state-of-the arts.
The training time of the hypernetwork (13 GPU hours) was lower than those of the ensemble methods and UNETRs although it was higher than those of the training of individual ResUNETs.  


\section{Discussion}
\label{sec:Discussion}

We proposed novel approaches to improve the segmentation probability map estimation by leveraging the network and hypernetwork ensembles with the varying Tversky loss~\cite{ha2016hypernetworks,hoopes2021hypermorph,salehi2017tversky,hansen1990neural,zhou2021ensemble}. 
The proposed methods successfully estimated the underlying segmentation probability maps that reflected the ambiguity for the challenging medical image segmentation task without sacrificing the binary segmentation performance compared to the state-of-the-arts. 
Although our approaches were only tested on the binary 3D image segmentation problem, our proposed approaches can be easily extended to wider applications. 

There are a few limitations of the proposed methods. The proposed hypernetwork architecture is implemented with the fully-connected dense layers~\cite{rumelhart1985learning} in the HyperConv block that inflated the size of the hypernetwork. 
This can be addressed by employing the efficient encoder designs~\cite{ronneberger2015u,kramer1991nonlinear,kingma2013auto}.
The quality of the estimated segmentation probability maps were not quantitatively evaluated due to the lack of multiple annotations per image in our dataset. The alternative approach can be the generalized energy distance~\cite{kohl2018probabilistic,lee2016stochastic,monteiro2020stochastic}. 
Also, it can be evaluated semi-quantitatively by the human expert rejection rate often used for an ambiguous segmentation problem~\cite{billot2020automated}. 
Lastly, our experiments did not cover the stochastic networks~\cite{monteiro2020stochastic,lee2016stochastic,kohl2018probabilistic}. Although they focused on the generation of the multiple segmentation label candidates and the estimation of the segmentation uncertainty, the segmentation probability map can be obtained by averaging the generated candidates at inference that can be comparable to ours. 

We believe our methods can contribute to many segmentation problems, not only limited to medical applications, that suffer from the inherent label ambiguity and the segmentation probability maps are highly desirable. 

\vspace{20pt}
\noindent \textbf{Acknowledgements }This research was supported by NIH NINDS MRI-GENIE:
R01NS086905, K23NS064052, R01NS082285, NIH NIBIB NAC P41EB015902 and NIH NINDS U19NS115388. 

{\small
\bibliographystyle{ieee_fullname}
\bibliography{bib}
}

\newpage

\appendix

\counterwithin{equation}{section}

\renewcommand{\figurename}{Figure S.}
\renewcommand{\tablename}{Table S.}
\setcounter{figure}{0}
\setcounter{table}{0}

\begin{table*}[t!]
\footnotesize
\begin{tabular}{lcccccccccc}
\toprule
                 & Conf. 1. & Conf. 2. & Conf. 3. & Conf. 4. & Conf. 5. & Conf. 6.                                                            & Conf. 7.                                                            & \begin{tabular}[c]{@{}c@{}}Conf. 8.\\ (Ours)\end{tabular}  & \begin{tabular}[c]{@{}c@{}}Conf. 9.\\ (Ours)\end{tabular}  & \begin{tabular}[c]{@{}c@{}}Conf. 10.\\ (Ours)\end{tabular} \\
\midrule
\begin{tabular}[c]{@{}l@{}}Network\\ Architecture\end{tabular}  & ResUNet  & ResUNet  & UNETR    & ResUNet  & UNETR    & ResUNet                                                             & UNETR                                                               & ResUNet                                                    & UNETR                                                      & \begin{tabular}[c]{@{}c@{}}Hyper-\\ ResUNet\end{tabular}   \\
Loss                                                            & Dice     & Dice CE  & Dice CE  & Dice CE  & Dice CE  & Dice CE                                                             & Dice CE                                                             & \begin{tabular}[c]{@{}c@{}}Varying\\ Tversky\end{tabular}  & \begin{tabular}[c]{@{}c@{}}Varying\\ Tversky\end{tabular}  & \begin{tabular}[c]{@{}c@{}}Varying\\ Tversky\end{tabular}  \\
Drop-Out                                                        & None     & None     & None     & 0.1      & 0.1      & 0.1                                                                 & 0.1                                                                 & None                                                       & None                                                       & None                                                       \\
No. Networks                                                    & 1        & 1        & 1        & 1        & 1        & 5                                                                   & 5                                                                   & 5                                                          & 5                                                          & 1                                                          \\
\begin{tabular}[c]{@{}l@{}}No. Trainable\\ Params\end{tabular}  & 4.8M     & 4.8M     & 92.7M    & 4.8M     & 92.7M    & 4.8M (x5)                                                           & 92.7M (x5)                                                          & 4.8M (x5)                                                  & 92.7M (x5)                                                 & 130M                                                       \\
No. Epochs                                                      & 2,000    & 2,000    & 2,000    & 2,000    & 2,000    & 2,000                                                               & 2,000                                                               & 2,000                                                      & 2,000                                                      & 3,000                                                      \\
\begin{tabular}[c]{@{}l@{}}Minibatch \\ Size\end{tabular}       & 8        & 8        & 8        & 8        & 8        & 8                                                                   & 8                                                                   & 8                                                          & 8                                                          & 8                                                          \\
\begin{tabular}[c]{@{}l@{}}Pre-Ensemble \\ Setting\end{tabular} & N/A      & N/A      & N/A      & N/A      & N/A      & \begin{tabular}[c]{@{}c@{}}Data subsets/\\ Random Init\end{tabular} & \begin{tabular}[c]{@{}c@{}}Data subsets/\\ Random Init\end{tabular} & \begin{tabular}[c]{@{}c@{}}Random\\ Init\end{tabular}      & \begin{tabular}[c]{@{}c@{}}Random\\ Init\end{tabular}      & N/A                                                        \\
\begin{tabular}[c]{@{}l@{}}Ensemble \\ Strategy\end{tabular}    & N/A      & N/A      & N/A      & N/A      & N/A      & \begin{tabular}[c]{@{}c@{}}Average\\ Ensemble\end{tabular}          & \begin{tabular}[c]{@{}c@{}}Average \\ Ensemble\end{tabular}         & \begin{tabular}[c]{@{}c@{}}Average\\ Ensemble\end{tabular} & \begin{tabular}[c]{@{}c@{}}Average\\ Ensemble\end{tabular} & \begin{tabular}[c]{@{}c@{}}Average\\ Ensemble\end{tabular} \\ 
\bottomrule
\end{tabular}
\vspace*{-2mm}
\caption{The details of network architectures and training configurations of the comparison and proposed methods.}
\label{supp_tab:details}
\end{table*}

\section{Network Architectures and Training Details}
\label{supp_sec:arch}

\noindent \textbf{Network Architecture and Optimization} Table S.~\ref{supp_tab:details} summarizes the network architecture and training details of the comparison and proposed methods. The number of epochs of the proposed hypernetwork architecture (Hyper-ResUNet) was set to 3,000 because the learning rate of the Adam optimizer was set to $1e^{-5}$ compared to the other methods with 2,000 epochs and $1e^{-4}$ learning rate. 
The kernel depths of the segmentation networks of all ResUNet-based configurations except the Hyper-ResUNet were 16, 32, 64, 128, and 256 for the five layers of encoders and decoders.
The kernel depths of the Hyper-ResUNet was set to 32, 32, 64, 64, and 128. 
For the UNETR architecture, the sizes of the input patch, embedding, multilayer perceptron (MLP) sublayers were set to 16, 768, and 3,072, respectively, following their original implementation. The number of heads of the UNETR was set to 12. 

\noindent \textbf{Data Augmentation} The same basic data augmentation strategies were applied to all configurations. 
The intensity range of each image $\bm{I}$ was scaled to [0, 1].
For training, we randomly adjusted the contrast of images at each minibatch with random $\gamma \in \mathbf{U}(0.5, 4.5)$,
\begin{equation}
    \bm{I}_{r. contr.} = ( \bm{I} - \min( \bm{I} ) )^{\gamma} ) ( \max( \bm{I} ) - \min( \bm{I} ) + \min( \bm{I} ).
\end{equation} 
Images and human annotation maps were randomly flipped by x-axis with 0.1 probability.
Images and human annotation maps were randomly cropped to 192x192x16 patches which showed empirically best results for all configurations compared to 64x64x8 and 256x256x32 (e.g., full images). 
For inference, the intensity range of an input image was scaled to [0,1]. 
At the inference stage, the input image was cropped sequentially to 192x192x16 patches and inferred by the sliding window technique with 80\% overlap. 

\section{Segmentation Uncertainty Estimation}
\label{supp_sec:uncertainty}

We can directly compute a segmentation uncertainty map from a segmentation probability map by the entropy,
\begin{equation}
    H( \bm{y} ) = -\sum_{i=1}^{N_L} ( p(\bm{y}_i ) + \epsilon )log (p(\bm{y}_i) + \epsilon ),
\label{supp_eq:entropy}
\end{equation}
where $N_L$ is the number of labels (e.g., two for binary segmentation), $p(\bm{y}_i)$ is the probability of the label to be the $i^{th}$ label (e.g., $\bm{y}_1 = 0$ and $\bm{y}_2 = 1$ for binary segmentation). 
$\epsilon$ is a small number for numerical stability.
For binary segmentation, $H(\bm{y})$ is in the range of approximately (0.0, 0.7).  

\begin{figure*}[htb!]
\centering
  \subfloat{ \includegraphics[width=170mm]{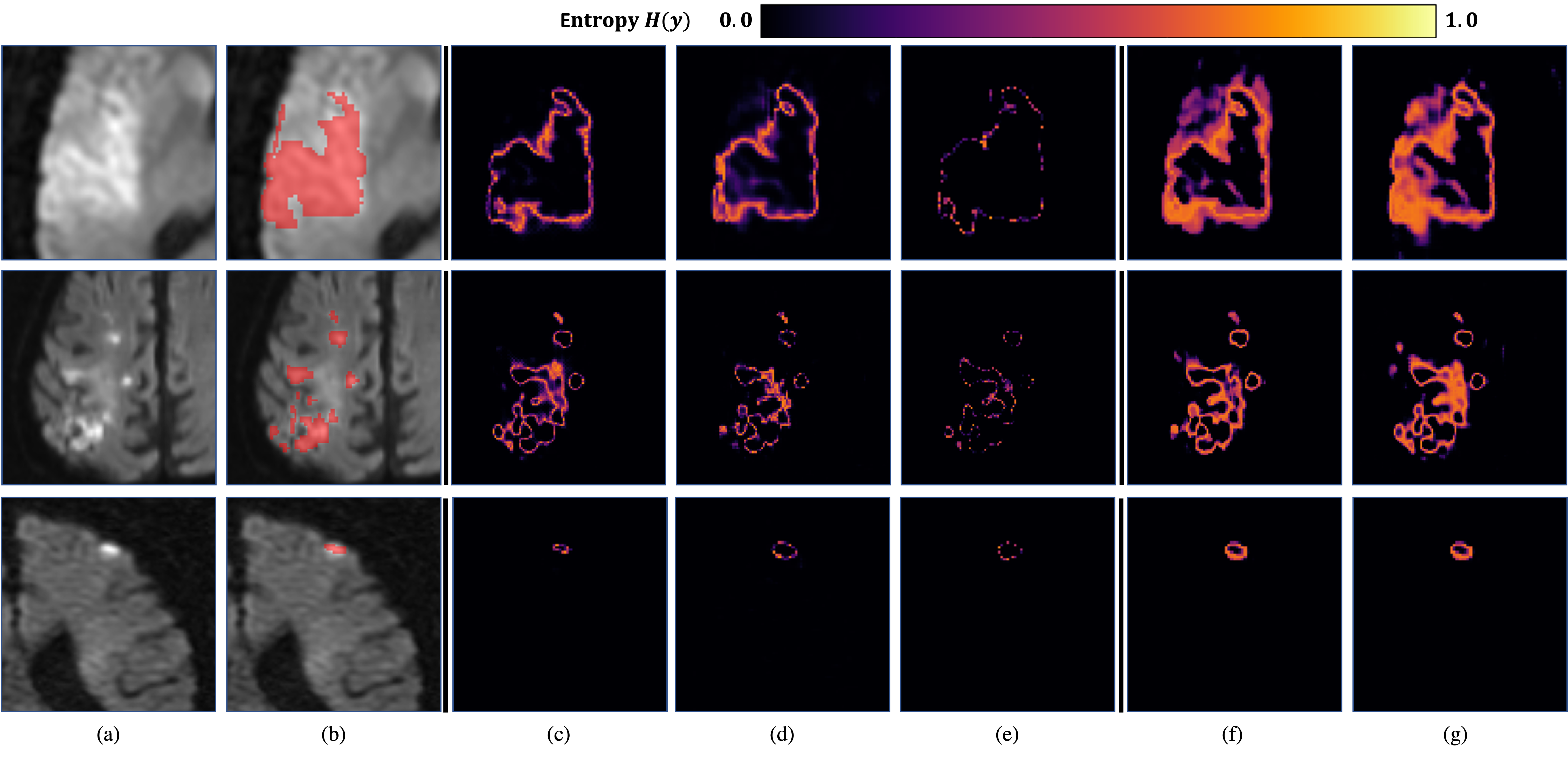} }
\vspace*{-3mm}
\caption{Examples of the estimated segmentation uncertainty maps. Zoomed and cropped input images and human annotations are shown in (a) and (b), respectively. (c-g) show the The uncertainty maps computed based on the segmentation probability maps. (c) The vanilla ResUNet, (d) UNETR with the drop-out rate of 0.1, (e) UNTER ensemble with 5-fold data subsets and random initialization, (f) our proposed UNETR ensemble with the varying Tversky loss, and (g) Hyper-ResUNet with the varying Tversky loss.}
\label{fig:Uncertainty}
\end{figure*}

Figure S.~\ref{fig:Uncertainty} shows the examples of the estimated uncertainty maps. 
We set $\epsilon=5e^{-5}$. 
The input images and human annotations were shown in Figure S.~\ref{fig:Uncertainty} (a) and (b), respectively. 
The uncertainty maps from the vanilla ResUNet with the Dice CE loss, the single UNTER with the drop-out rate of 0.1, and the UNETR ensemble with the 5-fold data subsets and random initialization are shown in Figure S.~\ref{fig:Uncertainty} (c-e), respectively.
Those from the proposed the UNETR ensemble with the varying Tversky loss and the Hyper-ResUNet are shown in Figure S.~\ref{fig:Uncertainty} (f) and (g), respectively.
The top two rows show the same examples with Figure~\ref{fig:ProbEstimates}. 
The areas that are ambiguous between acute stroke lesions, white matter hyperintensities, and possible image artifacts showed the high uncertainty. 
The same ambiguous areas showed the segmentation probability close to 0.5 in Figure~\ref{fig:ProbEstimates}. 
The bottom row of Figure S.~\ref{fig:Uncertainty} shows the example with the low ambiguity. 
The proposed methods (Figure S.~\ref{fig:Uncertainty} (f) and (g)) only captured the ambiguous area caused by lesion diffusion. 
It showed the proposed methods did not impose undesired additional uncertainty.

\section{Mapping Network}
\label{supp_sec:mapping}

\begin{table}[htb!]
\footnotesize
\begin{tabular}{ccccc}
\toprule
\begin{tabular}[c]{@{}c@{}}Hyper-\\ vector\\ Size\end{tabular} & \begin{tabular}[c]{@{}c@{}}No.\\ Mapping\\ Layers\end{tabular} & \begin{tabular}[c]{@{}c@{}}Primary Net\\ Kernel\\ Depths\end{tabular} & \begin{tabular}[c]{@{}c@{}}Total \\ No.\\ Params\end{tabular} & \begin{tabular}[c]{@{}c@{}}Training\\ Time\\ (GPU Hours)\end{tabular} \\ 
\midrule
None                                                           & None                                                           & {[}32,32,64,64,128{]}                                                & 6.0M                                                         &   x                                                                   \\
8                                                              & 2                                                              & {[}32,32,64,64,128{]}                                                & 18.1M                                                         & 19.4                                                                  \\
16                                                             & 3                                                              & {[}32,32,64,64,128{]}                                                & 34.1M                                                         & 19.4                                                                  \\
32                                                             & 4                                                              & {[}32,32,64,64,128{]}                                                & 66.2M                                                          & 19.8                                                                  \\
64                                                             & 5                                                              & {[}32,32,64,64,128{]}                                                 & 130M                                                          &       20.1                                                                \\ 
128                                                             & 6                                                              & {[}32,32,64,64,128{]}                                                 & 258M                                                          &       21.5                                                                \\
\bottomrule
\end{tabular}
\vspace*{-2mm}
\caption{The configurations of the hypernetworks with different mapping network sizes.}
\label{supp_tab:mapping}
\end{table}

\begin{figure}[htb!]
\centering
  \subfloat{\includegraphics[width=80mm]{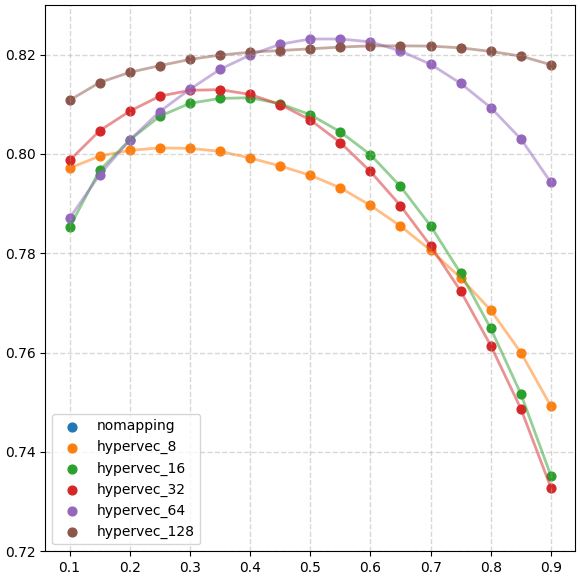} }
\vspace*{-3mm}
\caption{Dice scores with varying probability thresholds estimated by the hypernetworks with different mapping network sizes.}
\label{fig:ProbThres_Mapping}
\end{figure}

We investigate the effect of the different sizes of mapping networks and the mapped hypervectors in this section. The detailed configurations are summarized in Table S.~\ref{supp_tab:mapping}.
We fixed the configuration of the primary ResUNet to five-layer encoder and decoder with the kernel depths [32, 32, 64, 64, 128].
All hypernetworks were trained on NVIDIA Titan Xp (12GB GPU memory) with the batch size of 16 and 4,000 epochs.
Figure S.~\ref{fig:ProbThres_Mapping} shows the Dice scores of the segmentation label maps estimated by the hypernetworks with different segmentation probability thresholds.
The hypernetwork without the mapping network showed substantial decrease of the performance, 0.475 peak Dice score, and was not included in the plot.

We observed that the hypernetworks with larger mapping networks showed better Dice scores.
However, the change of the Dice scores with different segmentation probability thresholds decreased with the larger mapping network. 
We speculate that the hypernetwork might be overfit to data and human annotations when the size of the network increases. 
The performance improvement was also gradually saturated that the hypernetworks with the 4-layered and 5-layered mapping networks showed similar highest Dice scores. 

\section{Primary Network}
\label{supp_sec:primary}

\begin{table}[htb!]
\footnotesize
\begin{tabular}{ccccc}
\toprule
Conf. & \begin{tabular}[c]{@{}c@{}}Hypervector/\\ Mapping \\ Layers\end{tabular} & \begin{tabular}[c]{@{}c@{}}Primary Net\\ Kernel Depths\end{tabular} & \begin{tabular}[c]{@{}c@{}}Total\\ No. \\ Params\end{tabular} & \begin{tabular}[c]{@{}c@{}}Training\\ Time\\ (GPU Hours)\end{tabular} \\ 
\midrule
P. 1. & 32/4                                                                     & {[}32,64,128,256,512{]}                                         & 634M                                                          & 30.4$^*$                                                                 \\
P. 2. & 32/4                                                                     & {[}16,32,64,128,256{]}                                          & 158M                                                          & 19.1                                                                  \\
P. 3. & 32/4                                                                     & {[}32,32,64,64,128{]}                                           & 66.2M                                                         & 19.7                                                                  \\
P. 4. & 32/4                                                                     & {[}8,16,32,64,128{]}                                            & 39.7M                                                         & 16.9                                                                  \\
P. 5. & 32/4                                                                     & {[}16,16,32,32,64{]}                                            & 16.6M                                                         & 17.0                                                                  \\ 
\bottomrule
\end{tabular}
\vspace*{-2mm}
\caption{The configurations of the hypernetworks with different primary network sizes. All hypernetworks except Conf. P. 1. was trained on NVIDIA Titan Xp. ($^*$) Conf P. 1. was trained on NVIDIA QRTX 5000.}
\label{supp_tab:primary}
\end{table}

\begin{figure}[htb!]
\centering
  \subfloat{\includegraphics[width=80mm]{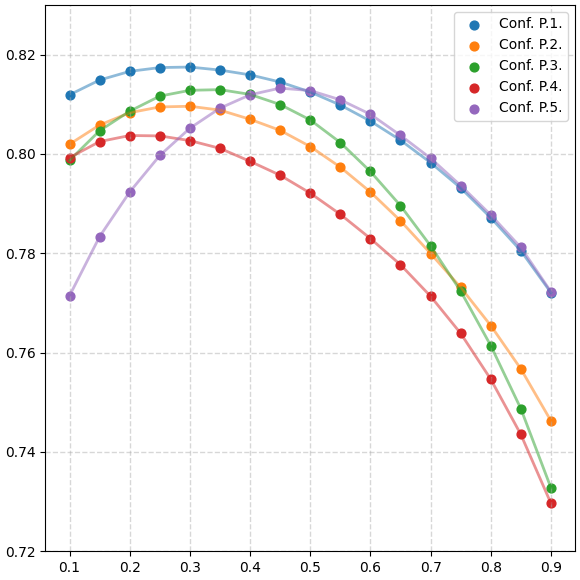} }
\vspace*{-3mm}
\caption{Dice scores with varying probability thresholds estimated by the hypernetworks with different primary network sizes.}
\label{fig:ProbThres_Primary}
\end{figure}

We analyze the effect of the size of the primary segmentation network (ResUNet) on the hypernetwork performance in this section. Table. S.~\ref{supp_tab:primary} summarizes the configurations of the hypernetworks with different primary network sizes. 
We fixed the hypervector size to 32 and the number of layers of the mapping network to four.
We fixed the batch size to 16 and the maximum epochs to 4,000. 
The hypernetworks were optimized by the Adam optimizer with the $1e^{-5}$ learning rate and $1e^{-3}$ weight decay.
Conf. P. 1. was trained on NVIDIA QRTX 5000 (16GB GPU memory) because of the large network size. 
The other configurations were trained on NVIDIA Titan Xp (12GB GPU memory).
Figure. S.~\ref{fig:ProbThres_Primary} shows the Dice scores of the hypernetworks with different primary network sizes.
The performance change with respect to different primary network sizes was not consistent: i.e., the larger primary network size did not always result in better performance although Conf. P. 1. with the largest primary network size showed the best performance. 
However, we observed the decrease of the change of Dice scores over different segmentation probability thresholds similar to the hypernetworks with the larger mapping networks. 
We speculate that a hypernetwork tends to overfit when an overall size increases while the effect of the mapping network may be greater. 




\end{document}